# The assembly integration and test activities for the new SOXS instrument at NTT

F. Biondi*[a], R. Claudi[a], L. Marafatto[a], J. Farinato[a], D. Magrin[a], R. Ragazzoni[a], S. Campana[b], P. Schipani[c], M. Aliverti[b], A. Baruffolo[a], S. Ben-Ami[d,e], A. Brucalassi[f,g], G. Capasso[c], R. Cosentino[h,l], F. D'Alessio[i], P. D'Avanzo[b], O. Hershko[d], H. Kuncarayakti[j,k], M. Munari[l], A. Rubin[d], S. Scuderi[l], F. Vitali[i], J. Achrén[m], J. Antonio Araiza-Duran[g,r], I. Arcavi[n], A. Bianco[b], E. Cappellaro[a], M. Colapietro[c], M. Della Valle[c], O. Diner[d], S. D'Orsi[c], D. Fantinel[a], J. Fynbo[o], A. Gal-Yam[d], M. Genoni[b], M. Hirvonen[p], J. Kotilainen[j,k], T. Kumar[k], M. Landoni[b], J. Lethi[p], G. Li Causi[i], S. Mattila[k], G. Pariani[b], G. Pignata[g,r], M. Rappaport[d], M. Riva[b], D. Ricci[a], B. Salasnich[a], R. Zanmar-Sanchez[l], S. Smartt[q], M. Turatto[a]

[a]INAF – Osservatorio Astronomico di Padova, Vicolo dell'Osservatorio 5, I-35122 Padova, Italy
[b]INAF – Osservatorio Astronomico di Brera, Via Bianchi 46 I-23807 Merate (LC), Italy
[c]INAF – Osservatorio Astronomico di Capodimonte, Salita Moiariello 16, I-80131, Napoli, Italy
[d]Weizmann Institute of Science, Herzl St 234, Rehovot, 7610001, Israel
[e]Harvard Smithsonian Center for Astrophysics, Cambridge, USA
[f]ESO, Karl Schwarzschild Strasse 2, D-85748, Garching bei München, Germany
[g]Universidad Andres Bello, Avda. Republica 252, Santiago, Chile
[h]FGG-INAF, TNG, Rambla J.A. Fernández Pérez 7, E-38712 Breña Baja (TF), Spain
[i]INAF – Osservatorio Astronomico di Roma, Via Frascati 33, I-00078 Monte Porzio Catone, Italy
[j]Finnish Centre for Astronomy with ESO (FINCA), FI-20014, University of Turku, Finland
[k]Tuorla Observatory, Department of Physics and Astronomy, FI-20014 University of Turku, Finland
[l]INAF – Osservatorio Astronomico di Catania, Via S. Sofia 78 30, I-95123 Catania, Italy
[m]Incident Angle Oy, Capsiankatu 4 A 29, FI-20320 Turku, Finland
[n]Tel Aviv University, Department of Astrophysics, 69978 Tel Aviv, Israel
[o]DARK Cosmological Center, Juliane Maries Vej 30, DK-2100 Copenhagen, Denmark
[p]ASRO (Aboa Space Research Oy) Tierankatu 4B, FI20520 Turku, Finland
[q]Astrophysics Research Centre, Queen's University Belfast, Belfast, County Antrim, BT7 1NN, UK
[r]Millennium Institute of Astrophysics (MAS)

## ABSTRACT

Son Of X-Shooter (SOXS) is the new instrument for the ESO 3.5 m New Technology Telescope (NTT) in La Silla site (Chile) devised for the spectroscopic follow-up of transient sources. SOXS is composed by two medium resolution spectrographs able to cover the 350-2000 nm interval. An Acquisition Camera will provide a light imaging capability in the visible band. We present the procedure foreseen for the Assembly, Integration and Test activities (AIT) of SOXS that will be carried out at sub-systems level at various consortium partner premises and at system level both in Europe and Chile.

*federico.biondi@inaf.it; phone +39 049 8293 534

# 1. INTRODUCTION

Son Of X-Shooter [1][2] (SOXS), Figure 1, will be a double spectrograph conceived and designed to be mounted at the Nasmyth-A focus of the ESO 3.5 m NTT in La Silla site (Chile). The instrument will represent a key facility for the spectroscopic follow-up of transient sources. The Italian National Institute of Astrophysics (INAF) is leading the project, collaborating with other Institutions around the world. At the moment, the instrument is in the Final Design Phase (with FDR in July 2018).

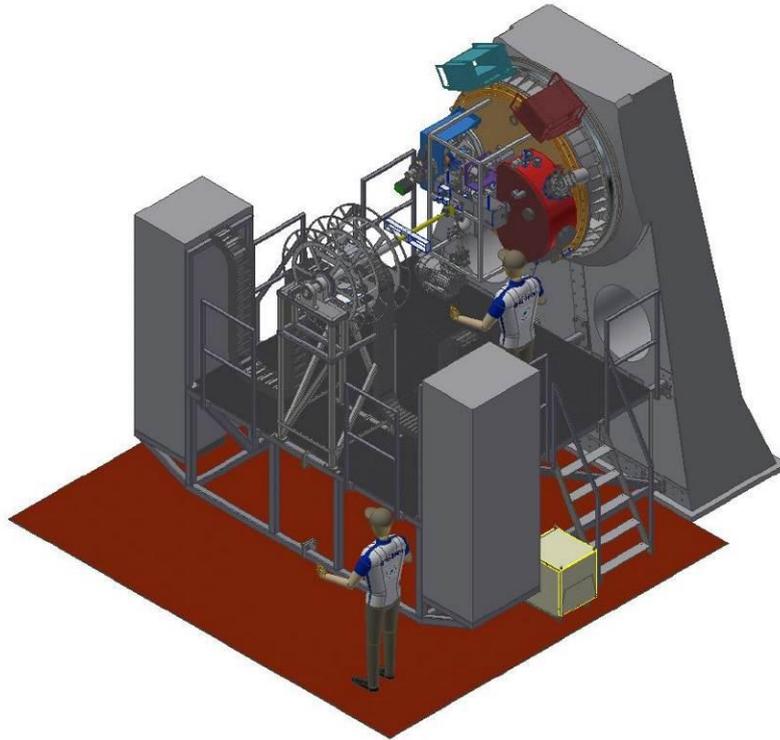

**Figure 1. A design of SOXS mounted at the NTT Nasmyth, with a new platform and co-rotator.**

SOXS design foresees two high-efficiency spectrographs with a resolution > 3500, able to cover the UV-VIS (350 – 850 nm) and NIR (800 – 2000 nm) bands. A light imaging capability in the visible band will be provided. The concept and design of SOXS is modular: the instrument is composed by five sub-systems in order to develop the AIT activities at the same time in different partner premises allowing to be on sky in early 2021. Each sub-system AIT phase is followed by an internal acceptance, the Assembly Readiness Review after which the AIT at system level can start. The AIT at system level is conduced twice: once at INAF Astronomical Observatory of Padova, in laboratory conditions, and then at telescope site in La Silla. The integration in Europe serves both as a full size test, allowing possible tuning of the procedures, and to define the mutual positions among sub-systems and between them and the telescope.

# 2. THE INSTRUMENT SUB-SYSTEMS

SOXS is designed to be modular: the main subsystems that constitute the instrument are the Common Path (CP) [3], UV-VIS Spectrograph (UVVIS-S) [4], [5], the NIR Spectrograph (NIR-S) [6], the Acquisition Camera (AC) [7] and the Calibration Unit (CU). In the following sections, we briefly describe them, referring the reader to [8], [9], [10], [11] for more detailed description of their mechanics, optics, electronics and software design, and we present the approach for the AIT procedure. Figure 2 shows the five sub-systems of SOXS.

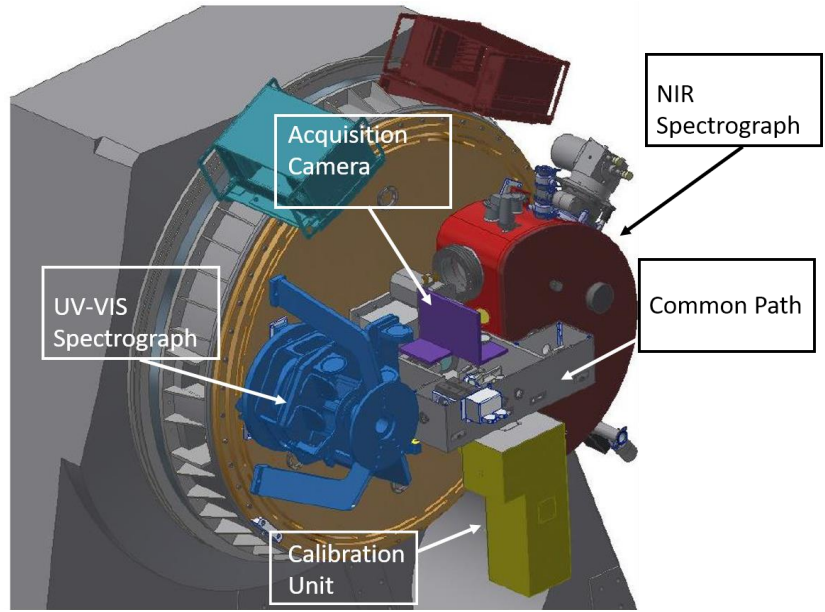

Figure 2. Design of SOXS sub-systems ingrated on the telescope interface flange.

## 2.1 Common Path

The CP (Figure 3) optically encloses the telescope Nasmyth focus, being the first subsystem that receives the F/11 telescope light-beam.

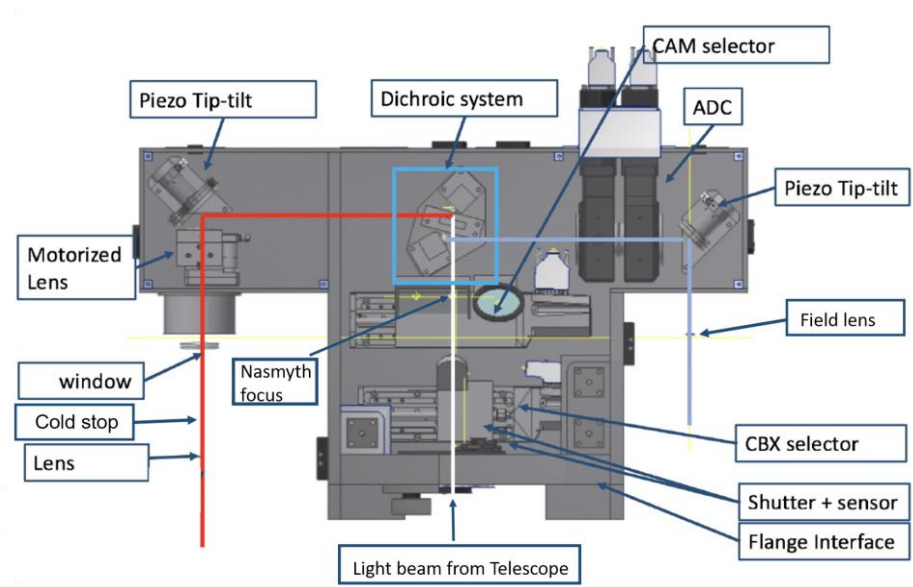

Figure 3. The layout of the Common Path.

The optical functionality of the CP consists in splitting the light in the two SOXS channels, the UV-VIS and the NIR, by means of a dichroic. Then it changes the F/# to 6.5 in both the two arms: in the UV-VIS channel, this is done by two doublets cemented on the prisms of an Atmospheric Dispersion Corrector; in the NIR channel the F/# is modified by a doublet that provides also the re-focuser mechanism for this arm. Two piezo-mirrors correct the position of the UV-VIS and NIR arms foci on the spectrograph slits. Since the CP represents also the interface among the spectrographs and the AC and CU, it carries also the opto-mechanical interfaces with these sub-systems, i.e. a slide selector for the AC and a slide selector for the CU.

The integration for the CP will be managed in Padova INAF premises. It involves first the characterization and the integration of the selection slide for the AC. This slide carries a rectangular flat mirror that can be placed in three different positions. In one position the full field can be reflected toward the AC; in the second position the mirror has a hole (15 arcsec) to provide both AC monitoring and spectroscopy; in the third position the mirror has a 0.5" pinhole. Since this pinhole coincides with the Nasmith focus, we use this reference to materialize the optical axis. Then we insert in the bench the dichroic, in nominal position by using a portable coordinate measuring machine (pCMM), touching the already characterized mechanical mount. We continue with the materialization of the optical path by using service pentaprisms, pinholes and targets. Then we integrate the mirrors, first positioning them with the pCMM and then checking if they change the already materialized optical path. The ADC (already assembled by the vendor) and the doublets are then aligned using their back reflected light as observable parameter for tip-tilt and decenter and looking at service detectors in the focal planes for tuning the position along the optical axis. The visible field lens, the NIR field lens, the NIR window and the NIR cold stop are integrated within the spectrograph AIT flux, while the CU slide is screwed in its nominal position. The tolerances for the placement of the elements of this sub-system range between 50 and 200 μm for centering and between 0.05 and 0.2 deg for tip-tilt.

All the electro-mechanical components are tested, a flexure test is foreseen. Tests on the transmissivity and on optical quality of the sub-system are due in order to accept the CP.

## 2.2 UV-VIS Spectrograph

The UVVIS-S works in the range 350 – 850 nm, which is splitted in four polychromatic beams of about 100 nm by three dichroics. Each quasi-order is diffracted by an ion-etched grating and then enters a three-element catadioptric camera that images them in a common detector. The optical layout of the spectrograph, of the camera and the two-level mechanical layout are shown in Figure 4. The first level of the system receives the light from the CP and includes a flat pickup mirror and an off-axis parabolic mirror used to feed the dichroics.

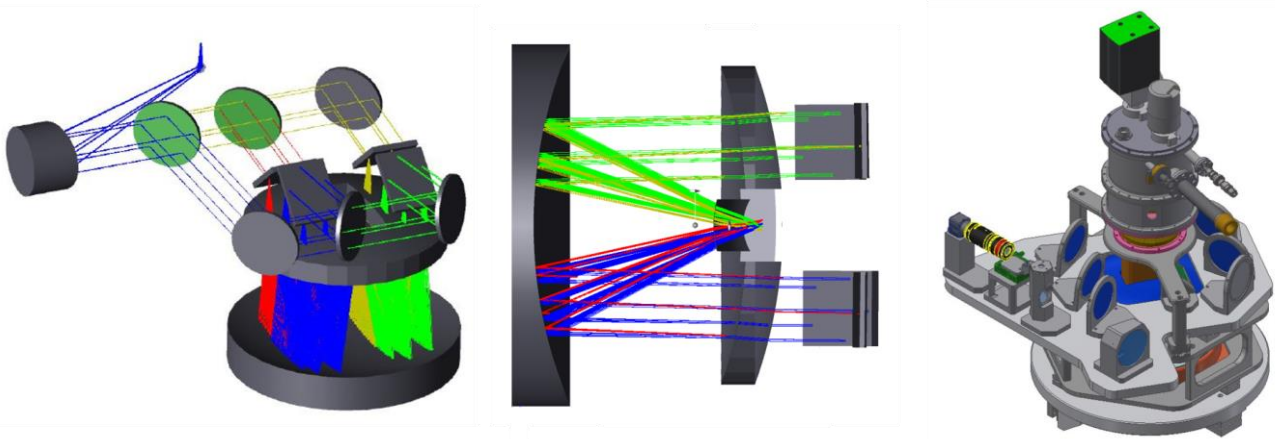

**Figure 4. Optical (left: spectrograph, center: camera) and mechanical (right) layout of UV-VIS Spectrograph.**

The dispersers and the camera are located at the second level of the sub-system. The camera is composed of three aspheric elements. The field flattener of the camera also serves as cryostat window, and is inserted through a rectangular aperture in the camera's corrector. The detector is located 4 mm behind the field flattener back surface. It is an e2V CCD44-82 CCD driven by the ESO NGC controller. The CCD will be cooled adopting a Continuous Flow Cryostat.

The UVVIS-S AIT will take place at Weizmann Institute in Rehovot, Israel. After the characterization of all the opto-mechanics by means of a CMM, these components are aligned mechanically placing them in nominal position, and then the optical alignment is checked, for this reason a CP simulator is foreseen. The two levels of the spectrograph, the feed plate and the camera, are mounted separately and then coupled each other with a kinematic mount system (3.1) in order to manage their mutual alignment. The estimated tolerances for this activity are of the order of 0.1 mm and 0.1 deg for centering and tip-tilt respectively. At the end, the already tested detector (@ FGG-INAF, TNG) detector cryogenic system (@ INAF-Catania premises) will be integrated after the delivering of the instrument in Padova.

A number of tests are foreseen for this sub-system. A full optical characterization in terms of throughput and optical quality is required. The functionality of the slit selector is checked. The system carries a detector that monitors the position of the slit and its adequate illumination, so a characterization of this element is necessary. A flexure test is foreseen in order to check the behavior of the spectral format during the rotation of the system held by a mechanical rotator.

## 2.3 NIR Spectrograph

The NIR-S is a cross-dispersed echelle based on the 4C concept, working in the range from 800 to 2000 nm. It is composed of a double pass collimator and a refractive camera, a grating disperser and a prism-based cross disperser. The design is shown in Figure 5.

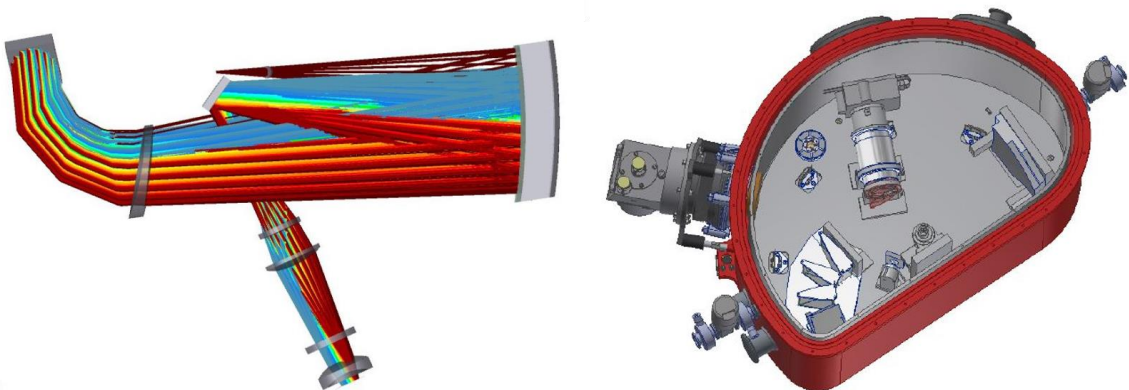

**Figure 5. Optical (left) and mechanical (right) layout of NIR Spectrograph.**

The spectrograph optomechanics will work at 150 °K in order to maintain adequate safety margins for the thermal radiation background, supposed to be well below the dark current of the array. The design includes an accurate baffling system as well as a (removable) thermal filter at 2 μm to cut off the longer wavelength radiation. The detector is a Teledyne H2RG array (to be tested in INAF-Roma premises). The vacuum and cryogenics system (to be tested in INAF-Catania premises) is based on a Closed Cycle Cryocooler.

The AIT of this sub-system will take place at INAF-Brera/Merate premises. The integration will adopt a mechanical metrology assisted alignment technique with a CMM. The integration will be performed in warm condition and an optical verification, by using a CP simulator, will certify the alignment to enable cooling down. At working temperature, the alignment is checked again and possible cold-warm iterations repeated. The tolerances are of the order of 50 μm in centering and 3 arcmin in tip-tilt of the camera elements and collimator mirror, doubled for the other elements.

The final verification of the sub-system foresees its optical characterization, through the execution of a series of calibration templates and the verification of the proper image quality, a functional test on the slit selector and preliminary flexure tests mounting the system on a rotator and recording the echellogram position and image quality during a rotation.

## 2.4 Acquisition Camera

The AC (Figure 6) is the sub-system that allows the acquisition of the target to be centered on the slits. In addition, it provides a facility for monitoring the co-alignment of the spectrographs, and it can also be used for imaging purposes in the visible band. At the NTT Nasmyth focal plane (so within the CP), a linear stage carries a mirror with three positions for different functions (2.1) and a pellicle beam splitter. The mirror re-directs the light from the telescope to the AC, the beam splitter from the spectrograph slits to the AC, in order to check their mutual alignment. The light bent by the mirror enters a camera, positioned on a re-focuser linear stage, then the optical path is tilted by a mirror, and encounters an eight positions filter wheel; the next camera, composed by two doublets and two singlets integrated in a tube, relays the Nasmyth focus on a detector with a F/# = 3.6.

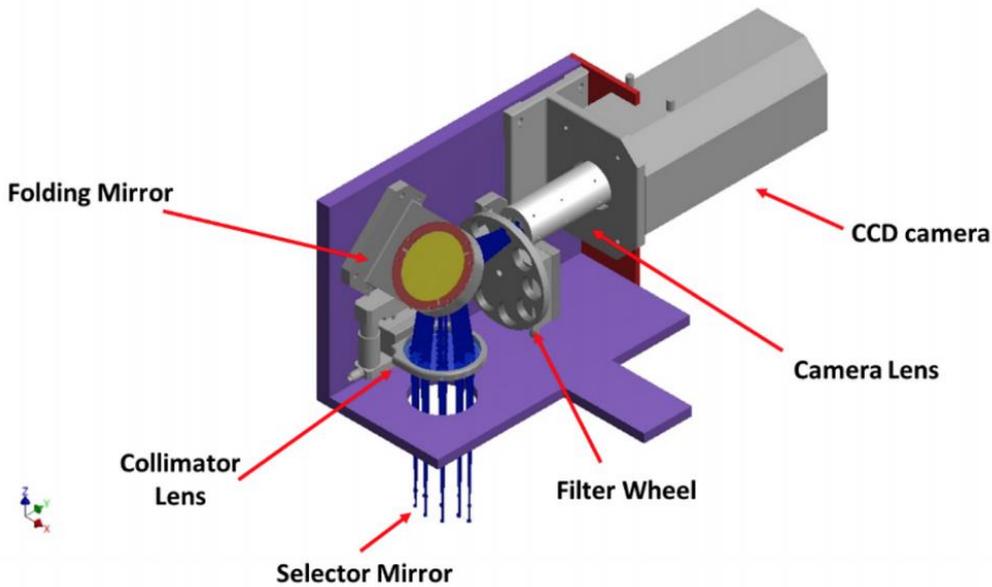

**Figure 6. The layout of the Acquisition Camera.**

The AC AIT will be managed in Santiago, Chile, at the Universidad Andres Bello. The two doublets and two singlets are inserted in their tubular mounting, centered and adjusted in tilt by means of three micrometric screws which have to be glued at the end of this alignment. The tubular mounting is linked to the CCD in its nominal position. Then, after the materialization of the optical path (with collimated source and mechanical references), the folding mirror is inserted and the re-focuser stage screwed in nominal position. At this point, the position of the lenses on the re-focuser is checked with the AC detector (decenter) and with a service CCD by looking at their back-reflected light (tip-tilt). By replacing the collimating source with a telescope simulator, one can control the focus on the detector and the possible vignetting of the filter wheel, after its integration. The tolerances for the optical element positions are 0.050 mm on decentering (0.040 mm on doublets) and 0.05 deg for tip-tilt. The final parameter for the alignment is the spot RMS radius with a value of 5.0 µm for the nominal design.

Besides the functional tests on the stages and on the cooling system for the detector, a full characterization of the CCD will be provided, considering typical test like the measurement of the of its quantum efficiency, read out noise, linearity, minimum and maximum integration time, size of usable image.

## 2.5 Calibration Unit

The CU (Figure 7) is the sub-system carrying the lamps that provide the spectra to calibrate the output of the spectrograph observations in terms of wavelength and flux. The light of the lamps, which covers the full wavelength range of the instrument, is collected and re-emitted by an integrating sphere. The exit pupil of the integrating sphere is imaged on the spectrographs slits, illuminating them uniformly, by means of a relay optic system. A mirror, mounted on a selection stage positioned in the CP, allows choosing the telescope or the CU input. The CU provides the possibility to create an artificial star for alignment operations using a pinhole mask than can be inserted in the optical path with a motorized stage.

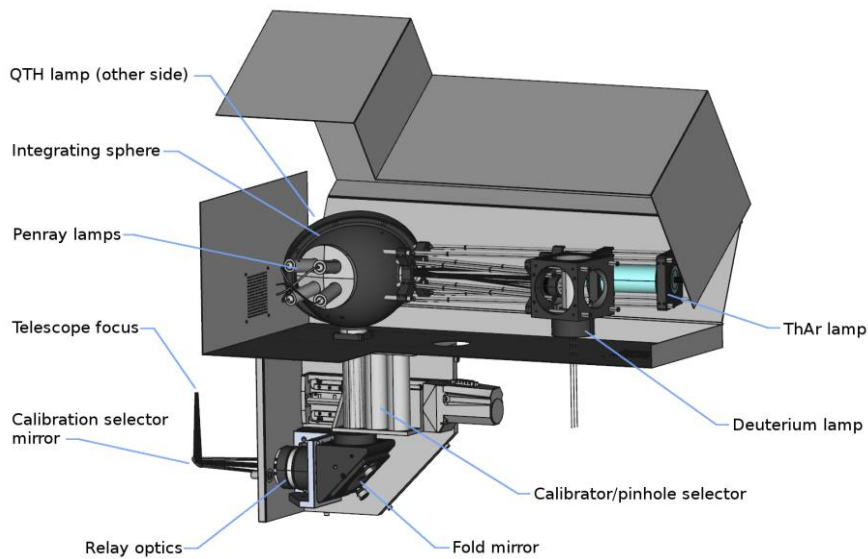

**Figure 7. The layout of the Calibration Unit.**

The integration and testing phase of the CU will be managed in Turku, Finland. Since the illumination patch is quite large with respect to the slit, the tolerances for the position of the optics within the mechanical bench of the CU are quite loose, making the integration process rather simple. The relay optics are aligned first by considering optical feedbacks for all the degrees of freedom. Then the integrating sphere, with lamps already mounted, is aligned to the relay optics by checking the uniformity of the illumination at the focal plane. The alignment of the pick-up mirror and its stage are considered a part of the CP AIT.

A number of tests are foreseen for the CU sub-system. The pinhole configuration is tested by a measurement of the PSF in focal plane and at different wavelenghts; a resolution test will be carried out. Every lamp will be tested before and after the beam passes through the relay optics to ensure the entire SOXS wavelength range is covered and the illumination patch is homogeneous and fit the expected size.

## 2.6 Assembly Readiness Review

Proceeding in parallel for the AIT sub-system phase allows speeding up the entire process of this short timescale project, but also implies a procedure to validate them before the shipment to INAF-Padova premises, where the integration at system level will take place. This phase is the Assembly Readiness Review, during which the responsible of the subsystems will carry out specific alignment tests, certifying the functionality and the performances of their instrument, in front of a panel constituted by the principal investigator, the project manager, the system engineer and the AIT manager of the project.

# 3. AIT AT SYSTEM LEVEL

We planned to perform the AIT at system level twice. First, we will mount the instrument at INAF premises in Padova for a full AIT test, a possible tuning of the procedures and a mechanical characterization of the sub-system components positions among them and with respect to the SOXS-Telescope Interface Flange. Then SOXS will be dismount, sent to Chile, re-assembled and tested again.

## 3.1 AIT in Europe

After the ARR, the sub-systems are delivered to Padova, and their alignment checked again to be ready to be integrated within the system as "stand-alone boxes". An optical simulator of the telescope is aligned to the rotation axis of a mechanical simulator of the NTT-Nasmyth bearing mechanism, which is provided by ESO. SOXS is interfaced to the simulator by means of a custom flange. All the sub-systems are linked together (CP with CU, CP with AC) or with the flange (CP with flange, UVVIS-S with flange, NIR-S with flange) by coupling the two halves of three kinematic mounts (KM) [12]. These mounts consist in three couples of steel plates: one hosts a sphere (which allows for 3 degrees of freedom), one a cylinder (2 degrees of freedom), one a washer (1 degree of freedom). If the mutual position of two sub-systems should be corrected, a proper shim is inserted between the KM and one of the two sub-systems or between the KM and the interface flange. A conservative evaluation for the repeatability of the KM system is 40 µm peak to valley.

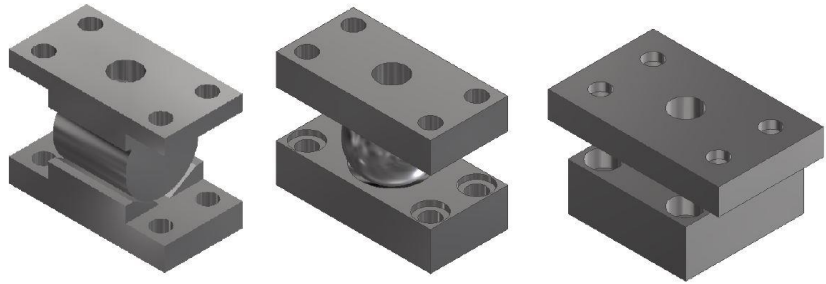

**Figure 8. Mechanical design of the three kinematic mounts used to link the sub-system among them and with the interface flange.**

The sub-systems carries mechanical references whose position is characterized with respect to the optical path, so by touching these references with a portable control measuring machine, we are able to place the various parts of the system in nominal position, then looking to an optical feedback for fine tune the alignment.

The CP is the first system to be linked to the flange. It is held through its ringbolts by an overhead crane, placed near a column, already fastened to the interface flange, and screwed to it in its nominal position. It is illuminated by the simulator and the artificial star focus should be in the center of the selection slide pinhole. In order to check the center and the tilt of the system with respect to the flange, we iteratively check the passage of the light through the pinhole and the footprint on the optical components. The exact position in the direction of the optical axis is provided by the image recorded on two CCDs, placed in the foci of the UV-VIS and NIR arms.

The CU is linked to the CP with the KM in its nominal position. The CU pinhole is inserted in the optical path simulating a star. We switch between the CU source and the telescope simulator source in order to co-align the CU box expecting the same observable.

The AC is linked to the CP through its KM. Using the telescope simulator, or the CU with the pinhole mode, we illuminate the selection slide in the CP with the pinhole position inserted. In this configuration, the light passing through the pinhole continues its path on the CP, while the marginal part of the 45° mirror reflects the light on the AC. Thus, the feedback for the alignment in centering is the position of the shadow of the pinhole on the detector of the AC. The tip-tilt and the focus would be adjust by looking to the optical quality of the focused image. Center and tip-tilt are then tuned by shimming the KM. The focus can be adjust moving the lens on the re-focuser linear stage, or by shimming.

We hold the UVVIS-S through its ringbolts by an overhead crane, place it near the flange and connect them with the three KM, whose position is already characterized at the end of the sub-system AIT. The fine tune of the alignment is done by looking to optical feedbaks. The position along the optical axis is checked by measuring the size of a focused image for each quasi-order, illuminating the instrument with the proper wavelength and the pinhole mask on the slit. The slit position (decenter of the spectropragh with respect to the CP) is checked looking directly toward it with a service camera mounted inside the spectrograph bench. The tip-tilt is corrected by looking to the radiance uniformity across each quasi-order when we illuminate the UVVIS-S with the CU flat field lamp.

We hold the NIR Spectrograph through its ringbolts by an overhead crane, put it near the flange and screw to it in its nominal position through the KM, whose position is already characterized at the end of the sub-system AIT. Also in this case an optical feedback is necessary in order to fine tune the mutual position between the Spectrograph and the CP. The decenter parameter is the beam position on the pinhole slit. The tip-tilt feedback is provided by a flux analysis illuminating the slit. We can check the position along the optical axis by considering the spectrum sharpness.

SOXS will be full functionally tested in laboratory conditions. The spectrograph output will be characterized using the CU. A flexure test is scheduled in order to provide a preliminary look up table to the CP piezo mirror system. Since the SOFI platform and co-rotator present now at the NTT Nasmyth have been re-designed, we plan to mount and test also the co-rotator (Figure 9).

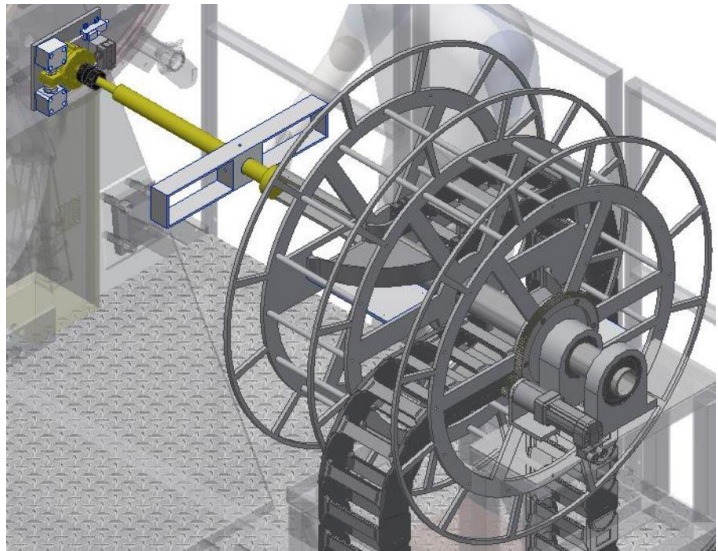

**Figure 9. The new corotator designed for SOXS.**

**3.2 AIT in Chile**

After the Preliminary Acceptance in Europe, the system is dismounted and sent to La Silla, where the process carried out in Italy is repeated again. Actually, given the high repeatability of the KM system, a simple coupling with the same shims should be enough to reach the foreseen alignment. The only component that, in general, could be linked differently is the flange with respect to the Nasmyth bearing. By characterizing this difference with a portable coordinate measuring machine, we are able to obtain the due correction to its placement, allowing being in nominal position with all the instrument. Of course, we are going to follow the step by step optical check during the all alignment.

After the whole integration, the system is characterized again in term of functionality and all the tests performed in Padova in laboratory conditions can be repeated with targets on sky. The spectrographs will be calibrated and characterized in terms of resolution, efficiency, spectral format, cross-calibration. A full flexure test is performed in order to provide a look up table for the tip tilt compensation operated by the piezo mirrors on the CP.

## 4. SUMMARY

We described the main phases of the AIT activities both at sub-system and system level for the SOXS instrument, highlighting its modularity terms of design and in terms of management of the procurement and integration. We show the different approaches to the integration, depending first on the tolerances and performances to be reached, but also on the different research groups' expertise and availability of facilities.

We stressed the importance of a double integration at system level, in Europe and in Chile, in order to check and improve the AIT procedure, to validate the instrument performances before deliver it to La Silla site, and to speed up the integration at the telescope.